\documentclass[aps,pra,nofootinbib]{revtex4-2}
\usepackage{epsfig}

\begin{document}
\date{}
\unitlength=1.00mm \special{em:linewidth 0.4pt}
\linethickness{0.4pt}
\thispagestyle{empty}
\newcommand{\be}{\begin{equation}}
\newcommand{\ee}{\end{equation}}
\newcommand{\ba}{\begin{eqnarray}}
\newcommand{\ea}{\end{eqnarray}}
\newcommand{\Gam}{\Gamma[\Phi]}
\newcommand{\Gamm}{\Gamma[\Phi,\rho]}
\newcommand{\si}{\sum_{i=1}^{N}}
\newcommand{\sij}{\sum_{i<j}}
\newcommand{\ri}{\textbf{r}_{i}}
\newcommand{\rj}{\textbf{r}_{j}}
\newcommand{\Phialpha}{\Phi(\textbf{r})}
\newcommand{\PhiN}{\Phi(\textbf{r}_{1},\textbf{r}_{2},\ldots,\textbf{r}_{N})}
\newcommand{\Phial}{\widehat{\Phi}_{\alpha}(\textbf{r},t)}
\newcommand{\Phigaher}{\widehat{\Phi}^{\dag}_{\gamma}(\textbf{r''},t'')}
\newcommand{\Phibe}{\widehat{\Phi}_{\beta}(\textbf{r'},t')}
\newcommand{\Phibeher}{\widehat{\Phi}^{\dag}_{\beta}(\textbf{r'},t')}
\newcommand{\Phideher}{\widehat{\Phi}^{\dag}_{\delta}(\textbf{r'''},t''')}

\title{Kohn-Sham approximation scheme \\for an interacting
Bose-condensed gas}

\author{Anna Okopi\'nska\\
Institute of Physics, Jan Kochanowski University,\\
Uniwersytecka 7, 25-406 Kielce, Poland\\and\\
Institute of Mathematics and Cryptology, Military University of Technology\\
Kaliskiego 2, 00-908 Warszawa, Poland\\
e-mail: anna.okopinska@wat.edu.pl }

\begin{abstract}
The grand canonical density functional theory for inhomogeneous
systems of interacting bosons is developed in the effective action
approach. The Legendre transform of the generating functional for
Green's functions is used to define the effective action as a
functional of both the particle density and the order parameter.
Expanding the thermal effective action in powers of the Planck
constant we obtain a systematic approximation scheme, which
practically implements the Kohn-Sham idea: the problem of
interacting bosons is reduced to a single-particle system in a
fictitious external potential. The Kohn-Sham potential, the density and
the order parameter have to be determined self-consistently in a
given order approximation.

\end{abstract}

\maketitle

\section{Introduction}
The Density Functional Theory (DFT), has become nowadays a method of
choice in quantum chemistry and solid state
physics~\cite{Parr,Drei}. The theory is based on the exact theorem
of Hohenberg and Kohn~\cite{HK} that a functional exists, by
minimization of which the density and other ground state properties
of the many-body system are completely determined. The practical
applications are successfully developed using the idea of Kohn and
Sham~\cite{KS} to replace the interacting many-electron problem by
the exactly equivalent problem of non-interacting particles moving
in an appropriately chosen external potential. Solving the
single-particle problem numerically is a standard task, all the
difficulties being transferred to the construction of the Kohn-Sham
potential. The rigorous definition of the density functional (DF), was given by
Levy~\cite{Levy} and Lieb~\cite{Lieb} in the constrained search
approach. Later, an extension to finite temperatures
has been discussed by Mermin in the grand canonical ensemble~\cite{Mer}. 
The constrained search approach does not provide, however, an explicit 
method to construct the DF; therefore various physically motivated 
approximate forms have been guessed and discussed in many works, both for 
solids and for molecules~\cite{Parr,Drei}. 

Several years ago, Fukuda et al.~\cite{Fukuda} provided a
new formulation of DFT, using generating functionals with an
external source $J(\textbf{r},t)$ linearly coupled to the local
composite operator $\widehat{\Phi}^{+}(\textbf{r})\widehat{\Phi}(\textbf{r})$. The
effective action, obtained as the Legendre transform of the generating
functional for connected Green's functions, has been used to define
the time-dependent DF in a way different from that developed by
Runge and Gross~\cite{tdep}. Using the path integral representation,
Fukuda et al.~\cite{Fukuda} were able to express the DF as a series
in powers of the interaction strength, formulating diagrammatic
rules for coefficients~\cite{Oku, Yoko}. 
Extending the formalism to finite temperature, Valiev and Fernando~\cite{Val} demonstrated that the
imaginary time effective action coincides with the Mermin grand
canonical DF. Moreover, they have shown that the approximation
scheme generated by the expansion in powers of the interaction
strength can be regarded as an implementation of the idea of Kohn
and Sham~\cite{KS}. The leading order approximation describes the
thermal equilibrium of non-interacting particles in an unknown
potential, which can is determined by higher-order corrections to
the effective action. An interesting modification of the
approximation scheme for DF in an effective field theory has
been proposed in nuclear physics with the effective parameter as a parameter
of expansion~\cite{Puglia}. The application of this scheme to dilute Fermi
system in a harmonic trap demonstrated the convergence of densities
and energies with increasing order calculations~\cite{Puglia,Furn}. 
An alternative scheme of gradually including interactions, motivated by 
renormalization group~\cite{Polonyi}, is successfully developed~\cite{AbInitio}.

In the case of bosons, the Hohenberg-Kohn theorem is valid, but as argued by Griffin~\cite{Grif}, the problem is complicated by the
phenomenon of Bose-Einstein condensation, which takes place
below the critical temperature. In the field-theoretic approach,
this is attributed~\cite{Bog} to the spontaneously broken symmetry
with the macroscopic wave-function
$\Phi(\textbf{r})=<\widehat{\Phi}(\textbf{r})>$ playing the role of
the order parameter, which determines the condensate density
$\rho_{c}(\textbf{r})=|\Phi(\textbf{r})|^{2}$. The proper extension
of the Hohenberg-Kohn theorem to bosonic fields makes it necessary
to consider a functional of both the particle density
$\rho(\textbf{r})$ and the order parameter $
\Phi(\textbf{r})$~\cite{Grif}. However, because of difficulties in
defining the Kohn-Sham reference system, the dependence on the order
parameter was never included in the practical applications of DFT
for bosons, and only the approximate functionals depending on the
total particle density have been discussed~\cite{Stringari}.

In this work, we show that the density functional for bosonic
systems can be conveniently defined as the effective action of
quantum field theory. We consider the connected generating
functional $W[j,J]$, which depends on two kinds of sources,
$J(\textbf{r},t)$ coupled to the composite density operator
$\widehat{\Phi}^{+}(\textbf{r})\widehat{\Phi}(\textbf{r})$, and
$j(\textbf{r},t)$ coupled to the elementary field
$\widehat{\Phi}(\textbf{r})$. The Legendre transform of $W[j,J]$
with respect to both sources defines the effective action as a
functional of the particle density $\rho(t,\textbf{r})$ and the
order parameter $\Phi(t,\textbf{r})$. We formulate a systematic
approximation scheme by expanding the effective action in powers of
the Planck constant. In the non-condensed phase, the expansion is
reduced to that in powers of the interaction strength, where the leading 
order Schr\"{o}dinger equation describes the non-interacting gas subjected to an unknown potential. In the
Bose-condensed phase, the leading order approximation is also of the
single-particle type, but given by the non-linear Gross-Pitaevskii
equation with an additional external potential. In both
cases the external potential is defined by higher-order
contributions to the effective action, and can be self-consistently
determined in the given order approximation. In this way a scheme
implementing the Kohn-Sham idea emerges naturally in this
approach and the many-body effects are taken into account in a systematic manner. The scheme is formulated for spatially inhomogeneous
systems, which is important in describing the properties of
Bose-condensed gases in magnetic traps.

The outline of the present work is as follows. The effective action
will be discussed in the Lagrangian approach, but first, in
Section~\ref{Ham}, we review briefly the grand canonical DFT in the
Hamiltonian approach in order to relate the two formulations.  In
Section~\ref{GF} generating functionals for Green's functions are
presented and the effective action is defined as a functional of the
order parameter and particle density. Expansion of the effective
action in powers of $\hbar$ is developed in Section~\ref{exp}.
Section~\ref{thermal} discusses the case of thermal equilibrium, and
the Kohn-Sham approximation scheme for interacting bosons is described in Section~\ref{appr}.
Our conclusions are summarized in Section~\ref{con}.

\section{DENSITY FUNCTIONAL IN THE CONSTRAINED SEARCH
APPROACH}\label{Ham} The quantum many-body system is usually
described by the second quantized Hamiltonian  \ba
\widehat{H}&=&\widehat{T}+\widehat{U}+\widehat{V}_{ext}=- \int
\!d^{3}r\frac{\hbar^{2}}{2m}\widehat{\Phi}^{\dag}(\textbf{r})
\nabla^2\widehat{\Phi}(\textbf{r}) \nonumber\\&+&\int \!d^{3}r
V_{ext}(\textbf{r})\widehat{\Phi}^{\dag}(\textbf{r})
\widehat{\Phi}(\textbf{r})+\frac{1}{2}\!\int \!d^{3}r\int \!d^{3}r'
\widehat{\Phi}^{\dag}(\textbf{r})\widehat{\Phi}^{\dag}(\textbf{r'})
U(\textbf{r},\textbf{r'})
\widehat{\Phi}(\textbf{r'})\widehat{\Phi}(\textbf{r}),\label{Hop}
\ea with the inter-particle interaction $U(\ri,\rj)$, and the
external potential $V_{ext}(\ri)$ characterizing the considered
system (the potential of nuclei for a molecule or a solid, the
potential of magnetic forces for a trapped atomic gas, etc.). This
form of the Hamiltonian applies both to fermions and bosons, the
different particle statistics is included by the appropriate
commutation relations for the field
operator $\widehat{\Phi}(\textbf{r})$. 
The rigorous definition of the DF, provided by the constrained
search approach~\cite{Levy,Lieb}, can be extended to the systems at
fixed temperature $T=\frac{1}{\beta}$~\cite{Parr}. In the grand
canonical ensemble, the states of the system are represented by Fock
space density operators: \be
\widehat{\Gamma}\!=\!\sum_{N}\sum_{i=1}^{\infty}
p_{N}^{i}|\Phi_{Ni}\rangle\langle \Phi_{Ni}|\ee with $p_{N}^{i}$
being the probability of finding the system in the $N-$particle
state $|\Phi_{Ni}\rangle$. Since the Hamiltonian~(\ref{Hop})
commutes with the number operator $\widehat{N}=\int
\!d^{3}r\widehat{\rho}(\textbf{r})=\int
\!d^{3}r\widehat{\Phi}^{\dag}(\textbf{r})\widehat{\Phi}(\textbf{r})$,
and the total number of particles $N$ is conserved, one introduces a
chemical potential $\mu$ with a value adjusted such that the average
number of particles would be equal to $N$. The grand canonical
functional of the state, defined as \be
\omega^{state}[\widehat{\Gamma}]\!=\!Tr\left\{\widehat{\Gamma}(\widehat{H}-\mu
\widehat{N}+\frac{1}{\beta}ln \widehat{\Gamma}
)\right\},\label{freen}\ee reaches a minimum for the equilibrium
state, $\widehat{\Gamma}=\widehat{\Gamma}_{eq}$, its value
determines the grand canonical potential of the system \ba
\omega(\beta,\mu)=\omega^{state}[\widehat{\Gamma}_{eq}]
=\inf_{\widehat{\Gamma}}Tr\left\{\widehat{\Gamma}(\widehat{H}-\mu
\widehat{N}+\frac{1}{\beta}ln \widehat{\Gamma}
)\right\}.\label{GC}\ea The search for the minimum can be split in
two steps. First, the constrained search is performed over all
states $\widehat{\Gamma}[\rho]$ with the expectation value of the
density operator equal to the prescribed function $\rho(\textbf{r})$
as defined by \be Tr\left[\widehat{\Gamma}\widehat{\rho}\right]=
\sum_{N}\sum_{i=1}^{\infty} p_{N}^{i}\langle
\Phi_{i}|\widehat{\rho}|\Phi_{i}\rangle=\rho(\textbf{r}),\label{rho}\ee
and later the obtained functional is minimized over all possible
$\rho(\textbf{r})$. This allows Eq.~\ref{GC} to be represented as
\ba
\omega(\beta,\mu)\!=\!\inf_{\rho(\textbf{r})}\inf_{\widehat{\Gamma}\rightarrow
\rho}tr\left\{\widehat{\Gamma}[\rho](\widehat{T}+\widehat{U}+\widehat{V}_{ext}\!-\!\mu
\widehat{N}+\frac{1}{\beta}ln \widehat{\Gamma}[\rho]
)\right\}\!=\!\inf_{\rho(\textbf{r})}\left[F[\rho]\!+\!\int d^3r
\rho(\textbf{r})V_{ext}(\textbf{r})\right],~\label{GCcs}\ea where
the universal functional  \be
F[\rho]\!=\!\inf_{\widehat{\Gamma}\rightarrow
\rho}Tr\left\{\widehat{\Gamma}[\rho](\widehat{T}+\widehat{U}-\mu
\widehat{N}+\frac{1}{\beta}ln \widehat{\Gamma}[\rho]
)\right\}\label{HKGC}\ee does not depend on external potential.
The functional \be \Omega[\rho]=F[\rho]+\int d^3r
V_{ext}(\textbf{r})\rho(\textbf{r})\label{GCfun}\ee provides a
rigorous construction of the grand canonical DF, introduced by
Mermin~\cite{Mer}. Eq.\ref{GCcs} clearly shows that $\Omega[\rho]$
determines the equilibrium density and the grand canonical potential
of the interacting system by the minimum principle. The infimum is
searched in the class of functions which may be obtained from a Fock
space density matrix by~(\ref{rho}). All the functions fulfilling
the conditions \be \rho(\textbf{r})\geq 0,~~\mbox{and}~\int d^3r
\left|\nabla \rho^{1/2}(\textbf{r})\right|^{2}<\infty,\label{Gil}\ee
belong to this class, since any function of this type, normalized to
$N$, can be obtained from an $N-$particle density
matrix~\cite{Drei}. It was observed by Lieb~\cite{Lieb} that
regarding the grand potential ~(\ref{freen}) as a functional of an
arbitrary one-particle potential $V(\textbf{r})$, \ba
\omega[V]\!=\!\inf_{\widehat{\Gamma}}tr\left\{\widehat{\Gamma}
(\widehat{T}+\widehat{U}+\widehat{V}\!-\!\mu
\widehat{N}+\frac{1}{\beta}ln \widehat{\Gamma}
)\right\},\label{pot}\ea the universal DF~(\ref{HKGC}) can be
represented by the Legendre transform \be
F[\rho]=\sup_{V(\textbf{r})}\left[\omega[V]-\int d^3r
V(\textbf{r})\rho(\textbf{r})\right],\label{Le}\ee where the maximum
is searched over all reasonable functions $V(\textbf{r})$ at fixed
$\rho(\textbf{r})$. Knowing $F[\rho]$, the Mermin functional
$\Omega[\rho]$ can be easily obtained via Eq.\ref{GCfun}. For our
purposes, it is more convenient to represent the thermal DF directly
in terms of the Legendre transformation \be
\Omega[\rho]=\sup_{J(\textbf{r})}\left[W[J]-\int d^3r
J(\textbf{r})\rho(\textbf{r})\right],\label{Leg}\ee where \be
W[J]=\omega[V]\!=\omega[V_{ext}+J]\ee is the grand
potential~(\ref{freen}) regarded as a functional of a new functional
variable $J(\textbf{r})=V(\textbf{r})-V_{ext}(\textbf{r})$. Observe
that $J(\textbf{r})$ plays a role of a fictitious external potential
which adds to the potential $V_{ext}(\textbf{r})$, which really
operates in the considered system. The formula~(\ref{Leg}) explains
the key idea of DFT by the Legendre concept of switching between
different independent variables: the dependence on the fictitious
potential $J(\textbf{r})$ is replaced by the dependence on the
density distribution $\rho(\textbf{r})$~\cite{Lieb,Nal}. The local
Legendre transform~(\ref{Leg}) is a functional generalization of the
transformation from the chemical potential $\mu$ to the number of
particles $N$. Unfortunately, neither (\ref{HKGC}) nor (\ref{Leg})
does present a useful way to calculate $\Omega[\rho]$ in practice. A
more suitable form can be obtained if the functional $W[J]$ has
appropriate differentiability properties and the supremum in
Eq.~\ref{Leg} occurs at \be \rho(\textbf{r})=\frac{\delta W} {\delta
J(\textbf{r})}.\label{dif}\ee With the solution of the above
equation expressed as a functional of the density, $J[\rho]$, the
Legendre transform is obtained just by substitution: \be
\Omega[\rho]=\left[ W[J]-\int d^3r
J(\textbf{r})\rho(\textbf{r})\right]_{J=J[\rho]}.\label{LegJ}\ee
Although not shown explicitly, it has to be borne in mind that the
Mermin DF depends on the chemical potential. The value of $\mu$ is
determined by the relation \be N=-\frac{\delta \Omega} {\delta
\mu},\label{chem}\ee which ensures the average number of particles
to be equal to $N$.

\section{GENERATING FUNCTIONALS FOR GREEN'S FUNCTIONS}
\label{GF} Full information on the quantum system requires the
knowledge of all Green's functions, and can be conveniently encoded
in generating functionals, which describe the system probed by
external classical sources. Here we consider the generating
functional in the form
 \ba Z[j,J]\!=\!\left<T e^{-\frac{i}{\hbar}\int
dt \left(\widehat{H}-\mu\int d^{3}r
\widehat{\Phi}^{\dag}(\textbf{r})\widehat{\Phi}(\textbf{r})+\int
d^{3}r\!j^{*}(t,\textbf{r})\widehat{\Phi}(\textbf{r})+\int d^{3}r
j(t,\textbf{r})\widehat{\Phi}^{\dag}(\textbf{r})+\int d^{3}r
\widehat{\Phi}^{\dag}(\textbf{r})J(t,\textbf{r})\widehat{\Phi}(\textbf{r})
\right)}\right>,~~\label{Z}\ea where the expectation value is taken
in the vacuum state, and $T$ denotes the time-ordering operator.
Besides the complex source $j(t,\textbf{r})$, linearly coupled to
the elementary quantum field $\widehat{\Phi}(\textbf{r})$, a real
source $J(t,\textbf{r})$, coupled to the density operator
$\widehat{\Phi}^{\dag}(\textbf{r})\widehat{\Phi}(\textbf{r})$, has
been introduced for more efficient probing of the system. The above
functional can be conveniently represented~\cite{Negele} as a path
integral 
\be Z[j,J]=\int\! D\Phi D\Phi^{*}\,e^{\frac{i}{\hbar} \int
dtd^{3}r
 \left[\emph{L}[\Phi]+\mu
\Phi^{*}(t,\textbf{r})\Phi(t,\textbf{r})-
\!j^{*}(t,\textbf{r})\Phi(t,\textbf{r})-
j(t,\textbf{r})\Phi^{*}(t,\textbf{r})-
\Phi^{*}(t,\textbf{r})J(t,\textbf{r})\Phi(t,\textbf{r})\right]},\label{ZL}
\ee where the Lagrangian density, derived from the
Hamiltonian~(\ref{Hop}), reads \ba
L[\Phi]&=&i\hbar\Phi^{*}(t,\textbf{r})\frac{\partial\Phi
(t,\textbf{r})}{\partial t }+
\frac{\hbar^{2}}{2m}\Phi^{*}(t,\textbf{r})\nabla^{2}\Phi(t,\textbf{r})
\nonumber\\&-&V_{ext}(\textbf{r})\Phi^{*}(t,\textbf{r})
\Phi(t,\textbf{r})-\frac{1}{2}\!\int \!d^{3}r'
\Phi^{*}(t,\textbf{r})
\Phi^{*}(t,\textbf{r'})U(\textbf{r},\textbf{r'})
\Phi(t,\textbf{r'})\Phi(t,\textbf{r}).~~~~~~\label{Lcl} \ea In the
general case of a time-dependent system, the path integral for the
generating functional~(\ref{ZL}) is defined within the
Schwinger-Keldysh formalism~\cite{SK} on the three-branch contour in
the complex-time plane
$\{(-\infty,+\infty),(+\infty,-\infty),(-\infty,-\infty+i\beta)\}$.
The boundary conditions on the fields are periodic in imaginary
time, with a period being the inverse temperature,  $\beta={1\over
T}$. Although almost all attention in this work is given to
equilibrium applications, we keep the formulation general as long as
possible, having in mind possible studies of time-dependent issues.
The generating functional for connected Green's functions, $W[j,J]$,
is defined by \be Z[j,J]=e^{\frac{i}{\hbar}W[j,J]}.\label{W}\ee The
background field in the presence of external sources can be obtained
as \ba \Phi(x)=\frac{\delta W}{\delta
j^{*}(x)}=<\widehat{\Phi}(x)>_{j,J}~~ \mbox{and}~~
\Phi^{*}(x)=\frac{\delta W}{\delta
j(x)}=<\widehat{\Phi}^{\dag}(x)>_{j,J},\label{bg}\ea \noindent and
the total density \ba n(x)=\frac{\delta W}{\delta
J(x)}=<\widehat{\Phi}^{\dag}(x)\widehat{\Phi}(x)>_{j,J}=<\widehat{\rho}(x)>_{j,J}=
\hbar\rho(x)+|\Phi(x)|^{2}\label{dens}\ea consists of the
uncondensed particles density $\rho(x)$ and the condensate density
$n_{cond}(x)=|\Phi(x)|^{2}$. Here and in the following $x$ stands
for $(t,\textbf{r})$.
 \noindent
The effective action for composite density operator is defined as
the double Legendre transform
 \be \Gamma[\Phi,\rho]=W[j,J]-\int\! \Phi^{*}(x)j(x)\,d x
-\int\! j^{*}(x)\Phi(x)-\int\! J(x)\left(\hbar
\rho(x)+|\Phi(x)|^{2}\right)\,d x\label{ea}\ee with the sources
$j(x)$ and $J(x)$ eliminated in favor of $\Phi(x)$ and $\rho(x)$
with the aid of Eqs.\ref{bg} and \ref{dens}. The above functional
contains full information on the system
 in terms of $\rho$ and  $\Phi$,
 corresponding to external sources $j$ and $J$. Due to Legendre
transform properties, the effective action fulfils \be \frac{\delta
\Gamma}{\delta \Phi(x)}=-j^{*}(x),~~\frac{\delta \Gamma}{\delta
\Phi^{*}(x)}=-j(x) ~~\mbox{and}~~\frac{\delta \Gamma}{\delta
\rho(x)}=-\hbar J(x).\label{derr}\ee The original system is
recovered by setting sources to zero, its states can be thus
determined by solving the stationarity conditions \be \frac{\delta
\Gamma}{\delta \Phi(x)}= \frac{\delta \Gamma}{\delta \Phi^{*}(x)}
=0\label{sta1}\ee and \be \frac{\delta \Gamma}{\delta
\rho(x)}=0.\label{sta2}\ee Since the interaction potential does not
depend on time, a time-independent solution can be found,
$\Phi_{eq}(x)=\Phi_{eq}(\textbf{r})$ and $
\rho_{eq}(x)=\rho_{eq}(\textbf{r})$, which corresponds to the
equilibrium state.

Let us observe that the conventionally used effective action \be
\Gamma[\Phi]=W[j,J=0]-\int\! \Phi^{*}(x)j(x)\,d x -\int\!
j^{*}(x)\Phi(x)\label{conv}\ee can be obtained as
$\Gamma[\Phi,\rho]$ at $J(x)=0$, or equivalently as.
$\Gamma[\Phi]=\Gamma[\Phi,\rho_{0}[\Phi]],$ where $\rho_{0}$ is a
solution of~(\ref{sta2}). Both the conventional effective action,
$\Gamma[\Phi]$, and the effective action for composite density
operator, $\Gamma[\Phi,\rho]$, contain full information on quantum
field theory, but considering $\Gamma[\Phi,\rho]$ as a functional of
two independent variables provides an easier access to some physical
observables. Both effective actions can be used to generate proper
vertices, which are the simplest, one-particle irreducible Green's
functions, directly related to the excitations of the system. Proper
vertices of elementary fields, defined trough differentiation of
$\Gamma[\Phi]$, can be also obtained as derivatives of
$\Gamma[\Phi,\rho]$ taken at the equilibrium values of the order
parameter and density. Especially useful is the second derivative
\be \Gamma(x,y)=\left(\begin{array}{cc}
 \Gamma_{\Phi\Phi^{*}}(x,y)&
\Gamma_{\Phi\Phi}(x,y)\\\Gamma_{\Phi^{*}\Phi^{*}}(x,y)
&\Gamma_{\Phi^{*}\Phi}(x,y)
\end{array}\right)= \left(\begin{array}{cc}
\left.\frac{\delta^{2} \Gamma}{\delta\Phi(x)
\delta\Phi^{*}(y)}\right|_{\Phi_{eq}\atop \rho_{eq}}&
\left.\frac{\delta^{2} \Gamma}{\delta\Phi^{*}(x)
\delta\Phi^{*}(y)}\right|_{\Phi_{eq}\atop \rho_{eq}}\\
                                      \left.\frac{\delta^{2} \Gamma}{\delta\Phi(x)
\delta\Phi(y)}\right|_{\Phi_{eq}\atop \rho_{eq}} &
\left.\frac{\delta^{2} \Gamma}{\delta\Phi^{*}(x)
\delta\Phi(y)}\right|_{\Phi_{eq}\atop \rho_{eq}}\\
                                   \end{array}\right),\label{iGf} \ee
which fulfils \be \int \Gamma(x,y)G(y,z)dz=-\delta (x,z),\ee where
the full propagator $G(x,y)$ is given by the connected Green's
function \ba G(x,y)= \left(\begin{array}{cc}
 G_{\Phi\Phi^{*}}(x,y)&
G_{\Phi\Phi}(x,y)\\G_{\Phi^{*}\Phi^{*}}(x,y) &G_{\Phi^{*}\Phi}(x,y)
\end{array}\right)= \left(\begin{array}{cc}
 \left.\frac{\delta^{2}
W}{\delta j(x) \delta j^{*}(y)}\right|_{j=J=0}&
\left.\frac{\delta^{2}
W}{\delta j^{*}(x) \delta j^{*}(y)}\right|_{j=J=0}\\
                                      \left.\frac{\delta^{2}
W}{\delta j(x) \delta j(y)}\right|_{j=J=0} & \left.\frac{\delta^{2}
W}{\delta j^{*}(x) \delta j(y)}\right|_{j=J=0}\\
                                   \end{array}\right).\ea Zero modes
of $\Gamma(x,y)$, corresponding to the poles of the propagator
$G(x,y)$, describe thus the one-particle excitations.

The functional $\Gamma[\Phi,\rho]$ offers an additional possibility
of taking functional derivatives with respect to the density, which
can be useful in studying collective excitations. For instance, the
density fluctuations are described by the two-point composite
vertex, given by the second derivative \be
\chi(x,y)=\left.\frac{\delta^{2} \Gamma}{\delta \rho(x)\delta
\rho(y) }\right|_{\Phi_{eq}\atop \rho_{eq}}. \label{iGfcom} \ee

\section{EXPANSION OF EFFECTIVE ACTION}\label{exp}

In the case of interacting particles, the effective action functionals cannot be
calculated exactly, so one resorts to approximations. It is
advantageous to formulate an approximation scheme for the effective
action functional, which makes it possible to generate consistent
sets of approximate Green's functions through functional
differentiation. A natural approximation scheme emerges if
$\Gamma[\Phi,\rho]$ can be represented as a series in powers of a
conveniently chosen parameter. Because of the implicit
definition~(\ref{ea}), the expansion of $\Gamma[\Phi,\rho]$ must be
obtained in three steps: expanding $Z[j,J]$ in powers of the chosen
parameter, deriving the expansion for $W[j,J]=\ln Z[j,J]$, and
performing the Legendre transform order by order in the chosen
parameter. Expansions of effective actions were obtained in this way for the cases when only one source is present. Expanding the conventional effective action $\Gamma[\Phi]$, being the Legendre transform of $W[j,J=0]$, in powers of the Planck constant, results in the
well-known loop expansion, represented by Feynman diagrams with 
$\Phi$-dependent propagator and vertices~\cite{Negele, Jackiw}. Expansion of
the density-dependent effective action $\Gamma[\rho]$, being the Legendre transform of $W[j=0,J]$ in powers of the interaction strength has been derived by Fukuda et al.~\cite
{Fukuda}. The diagrammatic representation of $\Gamma[\rho]$ has been established
in terms of the propagator, being related to $\rho(x)$ via an
implicit relation~\cite{Oku,Yoko}. Later, considering the effective action for composite operators $\widehat{\Phi}^2(\textbf{r})$ and $\widehat{\Phi}^4(\textbf{r})$ we have shown~\cite{AOIJMP} that by using
the Planck constant as a parameter of expansion, the Fukuda's
approach can be extended to the case when the effective action
depends on several functional variables. Now, we shall exploit this
idea to derive a diagrammatic representation of $\Gamma[\Phi,\rho]$, being the Legendre transform of $W[j]$ for non-relativistic system of interacting bosons.

The double Legendre transform for the effective action~(\ref{ea})
can be performed sequentially. First, we perform the Legendre
transform with respect to the source $j(x)$: \be
\Gamma_{1PI}[\Phi,J]=W[j,J]-\int\! \Phi^{*}(x)j(x)\,d x -\int\!
j^{*}(x)\Phi(x) d x,\label{first}\ee calculating the
integral~(\ref{ZL}) by the steepest-descent method. As stressed
in~\cite{Negele}, the steepest-descent method does not strictly
yield a semi-classical expansion in powers of $\hbar$, since the
Planck constant appears not only in the parameter $1\over \hbar$
multiplying the action in the exponent in~(\ref{ZL}) but also in the
Lagrangian~(\ref{Lcl}). Choosing the strategy, routinely used in
relativistic QFT, which consists in keeping the term $1\over \hbar$
in the exponent but setting $\hbar=1$ in the Lagrangian, we obtain
the steepest-descent expansion in the form \be
Z[j,J]=\sum_{k=0}^{\infty}\hbar^k Z^{(k)}[j,J].\ee This yields the
series representation of the connected generating functional \be
W[j,J]=\ln Z[j,J]=\sum_{k=0}^{\infty}\hbar^k W^{(k)}[j,J],\ee which
can be differentiated to obtain 
\be
\Phi[j,J]=\sum_{k=0}^{\infty}\hbar^k
\Phi^{(k)}[j,J]=\sum_{k=0}^{\infty}\hbar^k \frac{\delta
W^{(k)}[j,J]}{\delta j}.
\label{phi}
\ee 

The series for the background
field can be explicitly inverted order by order in $\hbar$ to the
form \be j[\Phi,J]=\sum_{k=0}^{\infty}\hbar^k
j^{(k)}[\Phi,J],\label{jot}\ee which enables $j(x)$ to be eliminated
in favor of $\Phi(x)$ in the Legendre transform~(\ref{first}) 
leading to the loop expansion formula
\be
\begin{picture}(139.00,80.00)
\put(0,70){{\large $\Gamma_{1PI}[\Phi,J]=\sum_{k=0}^{\infty}\hbar^k
\Gamma^{(k)}[\Phi,J]=S_{E}[\Phi]+\int J(x)|\Phi(x)|^2 dx+
\frac{\hbar}{2}TrLnG^{-1}$}}
\put(-5.00,42.00){\large $-\frac{3\hbar^2}{4}$}
\put(11.00,42.00){\circle{12.00}}
\put(17.00,42.00){\circle*{2.00}}
\put(23.00,42.00){\circle{12.00}} 
\put(42.00,11.00){\circle{12.00}}
\put(42.00,11.00){\oval(12.00,6.00)[]} 
\put(24.00,11.00){\large $-\frac{3\hbar^3}{4}$} 
\put(55.00,11.00){\large $-\frac{9}{4}\hbar^3$} 
\put(78.00,14.00){\circle{12.00}}
\put(15.00,10.89){\circle{12.00}} 
\put(-5.00,11.00){\large $-27\hbar^3$} 
\put(10.00,6.89){\line(1,2){5.00}}
\put(20.00,6.89){\line(-1,2){5.00}}
\put(85.00,6.00){\circle{8.00}} 
\put(71.00,6.00){\circle{8.00}}
\put(36.00,11.00){\circle*{2.00}} 
\put(48.00,11.00){\circle*{2.00}}
\put(74.00,9.00){\circle*{2.00}} 
\put(83.00,9.00){\circle*{2.00}}
\put(10.00,6.19){\circle*{2.00}} 
\put(20.00,6.89){\circle*{2.00}}
\put(15.00,16.89){\circle*{2.00}} 
\put(95.00,11.22){\large $+~O(\hbar^4)$} 
\put(30.00,41.00){{\large $-3\hbar^2$}}
\put(47.00,41.00){\circle{12.00}}
\put(41.00,41.00){\line(1,0){12.00}}
\put(41.00,41.00){\circle*{2.00}}
\put(53.00,41.00){\circle*{2.00}} 
\put(54.00,41.00){{\large $-81\hbar^3$}}
\put(75.00,41.00){\circle{12.00}}
\put(71.00,37.00){\line(0,1){8.00}}
\put(79.00,37.00){\line(0,1){8.00}}
\put(79.00,37.00){\circle*{2.00}}
\put(79.00,45.00){\circle*{2.00}}
\put(71.00,37.00){\circle*{2.00}}
\put(71.00,45.00){\circle*{2.00}}
\put(83.00,41.00){{\large $-27\hbar^3$}}
\put(103.00,41.00){\circle{12.00}}
\put(103.00,51.00){\circle{8.00}}
\put(103.00,47.00){\circle*{2.00}}
\put(96.00,41.00){\line(1,0){12.00}}
\put(96.50,41.00){\circle*{2.00}}
\put(109.50,41.00){\circle*{2.00}} 
\put(111.00,41.00){{\large $-54\hbar^3$}} 
\put(131.00,41.00){\circle{12.00}}
\put(131.00,41.00){\line(0,1){6.00}}
\put(126.00,36.00){\line(1,1){5.00}}
\put(136.00,36.00){\line(-1,1){5.00}}
\put(126.00,36.00){\circle*{2.00}}
\put(136.00,36.00){\circle*{2.00}}
\put(131.00,47.00){\circle*{2.00}}
\put(131.00,41.00){\circle*{2.00}}
\end{picture}
\label{Gamexp}
\ee 
with the line denoting the propagator functional $G_{J}(x,y)$, the inverse of
which is defined by 
\ba
G^{-1}_{J}(x,y)=~~~~~~~~~~~~~~~~~~~~~~~~~~~~~~~~~~~~~~~~~~~~~~~~~~~~~~~~~~~~~~~~~~~
~~~~~~~~~~~~~~~~~~~~~~~~~~~~~~~~~~~~~~~~~~~~~\nonumber\\
\left(\!\!\begin{array}{cc}
  \left(i\partial_{t}\!+\!\frac{\nabla^2}{2m}+V_{J}(x)\right)\!\delta(x\!-\!y)\!+2\Phi^{*}(x)U(x,y)\Phi(y)\!&
  \Phi(x)U(x,y)\Phi(y)\\
  \Phi^{*}(x)U(x,y)\Phi^{*}(y)
&\left(-i\partial_{t}\!+\!\frac{\nabla^2}{2m}+V_{J}(x)\right)\!\delta(x\!-\!y)\!+2\Phi^{*}(x)U(x,y)\Phi(y)
\end{array}\!\!\!\right),\label{prop}\ea
where the auxiliary potential $V_{J}(x)=-\mu+V_{ext}(x)+J(x)$. Dots
represent Hugenholtz vertices, which depend on the interaction
potential $U(\ri,\rj)$ and the background field $\Phi(x)$. The
diagrams in the above expansion should be interpreted according to
the rules of non-equilibrium theory on the three-branch contour in
the complex-time plane.\\

The next step consists in eliminating $J(x)$ in favor of $\rho(x)$,
while performing the second Legendre transform \ba
\Gamm=\Gamma_{1PI}[\Phi,J]-\int (\hbar\rho(x)+|\Phi(x)|^{2})J(x)
dx.\label{rem}\ea After substituting the loop expansion
(\ref{Gamexp}) into the relation \be |\Phi|^2+\hbar
\rho\!=\!\frac{\delta \Gamma_{1PI}}{\delta J}\!\ee one obtains the
power series representation of the density

\be
\begin{picture}(151.12,110.00)
\put(0.00,97.00){\makebox(0,0)[lc]{\large $\hbar\rho[\Phi,J]\!=\sum_{k=0}^{\infty}\hbar^k \rho^{(k)}[\Phi,J]\!=\hbar$}} 
\put(66.00,97.00){\circle{12.00}}
\put(73.00,97.00){\makebox(0,0)[lc]{\large $+\frac{\hbar^2}{2}$}}
\put(102.00,97.00){\circle{12.00}}
\put(89.00,97.00){\circle{12.00}}
\put(125.00,97.00){\circle{12.00}}
\put(110.00,97.00){\makebox(0,0)[lc]{\large $+\frac{\hbar^2}{3}$}} 
\put(95.50,97.00){\circle*{2.00}}
\put(119.00,97.00){\line(1,0){12.00}}
\put(60.00,97.00){\circle{2.00}}
\put(83.00,97.00){\circle{2.00}}
\put(125.00,91.00){\circle{2.00}}
\put(119.00,97.00){\circle*{2.00}}
\put(131.00,97.00){\circle*{2.00}}

\put(34.00,73.00){\makebox(0,0)[lc]{\large $+\frac{\hbar^3}{4}$}}
\put(45.00,61.00){\circle{2.00}}
\put(55.00,72.00){\circle{12.00}}
\put(48.00,64.00){\circle{8.00}}
\put(62.00,64.00){\circle{8.00}}
\put(51.00,67.00){\circle*{2.00}}
\put(59.00,67.00){\circle*{2.00}} 
\put(3.00,71.00){\large $+\frac{\hbar^3}{4}$} 
\put(23.00,72.00){\circle{12.00}}
\put(16.00,64.00){\circle{8.00}}
\put(30.00,64.00){\circle{8.00}}
\put(19.00,67.00){\circle*{2.00}}
\put(27.00,67.00){\circle*{2.00}} 
\put(72.00,71.00){\large $+\frac{\hbar^3}{2}$} 
\put(96.00,71.00){\circle{12.00}}
\put(90.00,71.00){\line(1,0){12.00}}
\put(88.00,65.00){\circle{8.25}}
\put(90.00,71.00){\circle*{2.00}}
\put(102.00,71.00){\circle*{2.00}}
\put(91.00,68.00){\circle*{2.00}}
\put(99.00,66.00){\circle{2.00}} 
\put(37.00,36.00){\large $+\frac{\hbar^3}{4}$}
\put(19.00,31.00){\circle{8.25}}
\put(112.00,71.00){\large $+\frac{\hbar^3}{2}$} 
\put(126.00,71.00){\line(1,0){12.00}}
\put(126.00,71.00){\circle*{2.00}}
\put(138.00,71.00){\circle*{2.00}}
\put(132.00,71.00){\circle{12.00}}
\put(124.00,64.00){\circle{8.25}}
\put(132.00,77.00){\circle{2.00}}
\put(127.00,67.00){\circle*{2.00}}
\put(23.00,78.00){\circle{2.00}}
\put(79.00,33.00){\circle*{2.00}}
\put(22.00,33.00){\circle*{2.00}}
\put(27.00,37.00){\circle{12.00}}
\put(21.00,37.00){\line(1,0){12.00}}
\put(21.00,37.00){\circle*{2.00}}
\put(33.00,37.00){\circle*{2.00}} 
\put(16.00,27.00){\circle{2.00}}
\put(89.00,33.00){\circle*{2.00}}
\put(89.00,33.00){\circle*{2.00}}
\put(84.00,43.00){\circle*{2.00}}
\put(118.00,37.00){\circle*{2.00}}
\put(106.00,37.00){\circle*{2.00}}
\put(52.00,33.00){\line(1,2){5.00}}
\put(62.00,33.00){\line(-1,2){5.00}} 
\put(3.00,37.00){\large $+\frac{\hbar^3}{4}$} 
\put(57.00,37.00){\circle{12.00}}
\put(57.00,31.00){\circle{2}} 
\put(57.00,44.00){\circle*{2.00}}
\put(62.00,33.00){\circle*{2.00}}
\put(52.00,33.00){\circle*{2.00}}
\put(68.00,37.00){\large $+\hbar^3$} 
\put(84.00,37.00){\circle{12.00}}
\put(93.00,37.00){\large $+\frac{\hbar^3}{6}$}
\put(112.00,37.00){\circle{12.00}}
\put(112.00,37.00){\oval(12.00,6.00)[]}
\put(79.00,33.00){\line(1,2){5.00}}
\put(89.00,33.00){\line(-1,2){5.00}}
\put(112.00,43.00){\circle{2.00}}
\put(79.00,40.00){\circle{2.00}}
\put(65.00,7.00){\large$+\frac{\hbar^3}{2}$}
\put(84.00,7.00){\circle{12.00}}
\put(80.00,2.00){\line(0,1){10.00}}
\put(88.00,2.00){\line(0,1){10.00}}
\put(80.00,2.00){\circle*{2.00}}
\put(88.00,2.00){\circle*{2.00}}
\put(88.00,12.00){\circle*{2.00}}
\put(80.00,12.00){\circle*{2.00}}
\put(77.50,8.00){\circle{2.00}}
\put(3.00,7.00){\large $+\frac{\hbar^3}{2}$}
\put(22.00,7.00){\circle{12.00}} 
\put(22.00,13.00){\circle{2.00}}
\put(18.00,2.00){\line(0,1){10.00}}
\put(26.00,2.00){\line(0,1){10.00}}
\put(18.00,2.00){\circle*{2.00}}
\put(26.00,2.00){\circle*{2.00}}
\put(26.00,12.00){\circle*{2.00}}
\put(18.00,12.00){\circle*{2.00}} 
\put(33.00,7.00){\large $+\frac{\hbar^3}{2}$}
\put(52.00,7.00){\circle{12.00}}
\put(51.80,7.50){\line(0,1){6.00}}
\put(47.00,3.00){\line(1,1){5.00}}
\put(57.00,3.00){\line(-1,1){5.00}}
\put(45.30,9.00){\circle{2.00}} 
\put(52.00,13.00){\circle*{2.00}}
\put(57.00,3.00){\circle*{2.00}}
\put(47.00,3.00){\circle*{2.00}}
\put(100.00,7.00){\large $+~O(\hbar^4)$}
\end{picture}
\label{rys2}
\ee 
which should be inverted to the form 
\be 
J[\Phi,\rho]=\sum_{k=0}^{\infty}\hbar^k J^{(k)}[\Phi,\rho].\label{Jot}
\ee  
In difference with the case of $j(x)$, the above inversion cannot be performed explicitly, since
the lowest order functional relation \be \rho^{(0)}(x)=
\frac{1}{2}tr G_{J}(x,x)\ee cannot be solved for $J(x)$. The best
thing one can do is to keep a definition of the functional
$J^{(0)}[\Phi,\rho]$ in an implicit form \be \rho(x)= \frac{1}{2}tr
G_{J^{(0)}}(x,x),\label{impl}\ee and to determine the higher-order
coefficients $J^{(k)}$ as functionals of $J^{(0)}$, which enables us
to perform the Legendre transform order by order in $\hbar$. For
simplicity, from here on we take the interaction potential to be
local \be U(\ri,\rj)=g \delta(\ri-\rj), \ee which is usually assumed
in describing the Bose condensed gas at very low energies, with
$g=\frac{4\pi\hbar^{2}a}{m}$ being related to the scattering length
$a$ \cite{Pethick}. In this case, the Hugenholtz vertices are reduced to the local
ones: the $3-$point vertices $-2g\Phi(x)$ and $-2g\Phi^{*}(x)$, and
the $4-$point vertex $-2g$, and the diagrammatic representation of
the effective action is obtained in the form

\be 
\begin{picture}(125.00,100.00)
\put(-6.00,90.00){\large $\Gamma[\Phi,\rho]= S[\Phi]+\hbar\int \rho(x)J^{(0)}(x)dx+\frac{\hbar}{2}TrLnG^{-1} $}
\put(0.00,67.00){\large $+\frac{3\hbar^2}{2}g\int\rho^2(x)dx$}
\put(34.00,67.00){\large $-3\hbar^2$}
\put(52.00,67.00){\circle{12.00}}
\put(46.00,67.00){\line(1,0){12.00}}
\put(46.00,67.00){\circle*{2.00}}
\put(58.00,67.00){\circle*{2.00}}
\put(60.00,67.00){\large$-81\hbar^3$} 
\put(82.00,67.00){\circle{12.00}}
\put(78.00,63.00){\line(0,1){8.00}}
\put(78.00,63.00){\circle*{2.00}}
\put(86.00,63.00){\circle*{2.00}}
\put(78.00,71.00){\circle*{2.00}}
\put(86.00,71.00){\circle*{2.00}}
\put(86.00,63.00){\line(0,1){8.00}} 
\put(90.00,67.00){\large $-54\hbar^3$}
\put(111.00,67.00){\circle{12.00}}
\put(111.00,67.00){\line(0,1){6.00}}
\put(111.00,67.00){\line(1,-1){5.00}}
\put(106.50,62.00){\line(1,1){5.00}}
\put(106.50,62.00){\circle*{2.00}}
\put(111.00,73.00){\circle*{2.00}}
\put(115.50,62.00){\circle*{2.00}}
\put(111.00,67.00){\circle*{2.00}}
\put(90.00,37.00){\circle*{2.00}} 
\put(102.00,37.00){\circle*{2.00}}
\put(96.00,37.00){\circle{12.00}} 
\put(78.00,37.00){\large $-\frac{3\hbar^3}{4}$} 
\put(96.00,37.00){\oval(12.00,6.00)[]}
\put(71.00,42.00){\circle*{2.00}} 
\put(71.00,30.00){\circle*{2.00}}
\put(52.00,42.00){\circle*{2.00}} 
\put(52.00,30.00){\circle*{2.00}}
\put(29.00,32.00){\circle*{2.00}} 
\put(19.00,32.00){\circle*{2.00}}
\put(24.00,42.00){\circle*{2.00}} 
\put(2.00,36.00){\large $-27\hbar^3$} 
\put(24.00,36.00){\circle{12.00}}
\put(19.00,32.00){\line(1,2){5.00}}
\put(29.00,32.00){\line(-1,2){5.00}} 
\put(31.00,36.00){\large $+81\hbar^3$} 
\put(71.00,36.00){\circle{12.00}}
\put(65.00,37.00){\line(-1,0){7.00}}
\put(58.00,35.00){\line(1,0){7.00}} 
\put(65.00,36.00){\circle{2.00}}
\put(58.00,36.00){\circle{2.00}} 
\put(60.00,33.00){\line(1,2){3.00}}
\put(71.00,30.00){\line(0,1){12.00}}
\put(52.00,36.00){\circle{12.00}}
\put(52.00,30.00){\line(0,1){12.00}}
\put(105.00,36.00){\large $+~O(\hbar^4)$} 
\put(-16.00,11.00){\Large where the inverse composite propagator} 
\put(82.00,14.00){\line(-1,0){7.00}}
\put(75.00,12.00){\line(1,0){7.00}}
\put(77.00,10.00){\line(1,2){3.00}}
\put(82.00,13.00){\circle{2.00}} 
\put(75.00,13.00){\circle{2.00}}
\put(85.00,11.00){\large $=\chi(x,y)=-\frac{\partial J^{(0)}(x)}{\partial \rho(y)}.$ }
\end{picture}
\label{eafull}
\ee 
The line denotes the propagator $G_{J}(x,y)$~(\ref{prop}) taken at
$J=J_{0}[\Phi,\rho]$, which is implicitly given by~(\ref{impl}). In fact, the implicit form of this relation is the advantage of the
method of composite operators, because more information is included
in the lowest order. It can be observed that for vanishing
background field, $\Phi=0$, the loop expansion~(\ref{eafull}) would
be reduced to an expansion of $\Gamma[\rho]$ in powers of the
interaction strength, similar to that obtained by Fukuda for
fermions~\cite{Fukuda}.

The lowest order of the time-dependent effective action is the
classical action \be \Gamma^{(0)}[\Phi,\rho]=S[\Phi]=\int dt
\!d^{3}r \left[\Phi^{*}(t,\textbf{r})\left(i\frac{\partial}{\partial
t }+ \frac{\nabla^2}{2m}-V_{ext}(\textbf{r})+\mu\right)\Phi
(t,\textbf{r})-\frac{g}{2}\!
\left|\Phi(t,\textbf{r})\right|^{4}\right],\label{Om0} \ee which
does not depend on $\rho$, and trivially fulfills~(\ref{sta2}). In
this approximation, the stationarity equation~(\ref{sta1}) yields
the time-dependent Gross-Pitaevskii equation \be \frac{\delta
\Gamma^{(0)}}{\delta
\Phi^{*}(t,\mathbf{r})}=\left(i\frac{\partial}{\partial t }+
\frac{\nabla^2}{2m}-V_{ext}(\textbf{r})+\mu-g\!
\left|\Phi(t,\textbf{r})\right|^{2}\right)\Phi
(t,\textbf{r})=0.\label{GP} \ee In order to determine corrections to
the above equation and other time-dependent characteristics, it
would be necessary to calculate higher-order diagrams of
$\Gamma[\Phi,\rho]$ by means of Schwinger-Keldysh rules. We will not
develop this point here, and in the following we restrict our
discussion to the equilibrium case, discussing a systematic
approximation scheme for the thermal DF.

\section{THERMAL DENSITY FUNCTIONAL}\label{thermal}

In the case of thermal equilibrium, the path integral formalism
becomes greatly simplified, since only the branch along the
imaginary axis on the Schwinger-Keldysh contour matters. Changing to
imaginary time $\tau=it$ reduces the grand canonical generating
functional to the Matsubara integral, \ba Z[j,J]=\int\! D\Phi
D\Phi^{*}\,e^{-\frac{1}{\hbar}\int dx \left[L_{E}[\Phi,\Phi^{*}]-\mu
\Phi^{*}(x)\Phi(x)+\!j^{*}(x)\Phi(x)+\!
j(x)\Phi^{*}(x)+\Phi^{*}(x)J(x)\Phi(x)\right]},\label{part}\ea where
$x$ stands for $(\tau,\textbf{r})$, and the integral over $\tau$ is
taken on the interval $(0,\beta)$, as the functions are periodic in
$\tau$. The Wick's rotated Lagrangian density takes a form \ba
L_{E}[\Phi]&=&\Phi^{*}(\tau,\textbf{r})\left(\hbar\frac{\partial}{\partial\tau}
- \frac{\hbar^{2}}{2m}\nabla^2+V_{ext}(\textbf{r})\right)\Phi
(\tau,\textbf{r})\nonumber\\&+&\frac{1}{2}\!\int \!d^{3}r'
\left(\Phi^{*}(\tau,\textbf{r})
\Phi^{*}(\tau,\textbf{r'}),\tau)U(\textbf{r},\textbf{r'})
\Phi(\tau,\textbf{r'})\Phi(\tau,\textbf{r})\right).\label{LEu} \ea
For studying the equilibrium properties of the system, it is
sufficient to consider time-independent generating functionals. The
functional \be
w[j,J]=-\left.\frac{1}{\beta}W[j,J]\right|_{j=j(\textbf{r})\atop
J=J(\textbf{r})}\ee represents the grand canonical potential of the
system being probed by the time-independent sources $j(\textbf{r})$
and $J(\textbf{r})$. In this case, the background field and density,
given respectively by \be \Phi(\textbf{r})=\frac{\delta w}{\delta
j^{*}(\textbf{r})},~~\Phi^{*}(\textbf{r})=\frac{\delta w}{\delta
j(\textbf{r})},~~ \mbox{and}~~ \frac{\delta w}{\delta
J(\textbf{r})}=\hbar\rho(\textbf{r})+|\Phi(\textbf{r})|^{2},\label{stasta}\ee
are also time-independent, and the effective action can be used to
define the thermal density functional \ba
\Omega[\Phi,\rho]\!=\!-\!\left.\frac{1}{\beta}\Gamma[\Phi,\rho]\right|_{\!\Phi(\textbf{r})\atop
\!\rho(\textbf{r})}\!=\!w[j,J]\!-\!\int
d^{3}r\Phi^{*}(\textbf{r})j(\textbf{r})\!-\!\int
d^{3}r\Phi(\textbf{r})j^{*}(\textbf{r})\!-\!\int\! J(x)\left(\hbar
\rho(x)+\!|\Phi(x)|^{2}\right).~~\label{tindepEA}\ea Since the
functional $w[j,J]$ is strictly concave in both variables, \be
w[j,\alpha J+(1-\alpha)J']>\alpha
w[j,J]+(1-\alpha)w[j,J'],~~~\mbox{for}~~~0<\alpha<1~~~\mbox{and}~~~
J\neq J'\ee \be w[\alpha j+(1-\alpha)j',J]>\alpha
w[j,J]+(1-\alpha)w[j',J],~~~\mbox{for}~~~0<\alpha<1~~~\mbox{and}~~~
j\neq j',\ee the transformation $j,J\rightarrow \Phi,\rho$, given
by~(\ref{stasta}), is bijective, i.e. for any $\Phi(\textbf{r})$ and
$\rho(\textbf{r})$ the corresponding $j(\textbf{r})$ and
$J(\textbf{r})$ exist. Therefore, the thermal density functional
$\Omega[\Phi,\rho]$ is strictly convex and provides an appropriate
extension of the Mermin functional~(\ref{LegJ}) to non-vanishing
background fields. The equilibrium values of the order parameter,
$\Phi_{eq}(\textbf{r})$, and density, $\rho_{eq}(\textbf{r})$, are
determined by the minimum principle \be \frac{\delta \Omega}{\delta
\Phi(\textbf{r})}=\frac{\delta \Omega}{\delta
\Phi^{*}(\textbf{r})}=0~~ \mbox{and}~~\frac{\delta \Omega}{\delta
\rho(\textbf{r})}=0,\label{staGam}\ee and the grand canonical
potential can be obtained as
$\omega(\beta,\mu)=\Omega[\Phi_{eq},\rho_{eq}]$. The value of the
chemical potential is fixed by the additional condition \be
\frac{\delta \Omega}{\delta \mu}= -N,\label{che}\ee which guarantees
that the average number of particles is equal to $N$. One can notice
that in the limit of $\beta \rightarrow \infty$, $\Omega[\Phi,\rho]$
approaches the zero-temperature DF, which describes the ground state
properties of the bosonic system.

Expansion of $\Omega[\Phi,\rho]$ in powers of $\hbar$ is obtained
from Eq.\ref{eafull} by replacing the Feynman rules by those of the
imaginary-time formalism at fixed temperature $T=\frac{1}{\beta}$
and chemical potential $\mu$. In the frequency-coordinate
representation, the thermal propagator reads
 \be
\mathcal{G}^{-1}_{J^{(0)}}\!(\omega_{n},\textbf{r},\textbf{r'})\!=\!\!\left(\!\!\begin{array}{cc}
  -i\omega_{n}-\!\frac{\nabla^2}{2m}\!+\!2g
|\Phi(\textbf{r})|^2\!+V_{J^{(0)}}(\textbf{r})& g
\Phi^{2}(\textbf{r})\\
g\Phi^{*2}(\textbf{r}) &i\omega_{n}-\!\frac{\nabla^2}{2m}\!+\!2g
|\Phi(\textbf{r})|^2\!+\!V_{J^{(0)}}(\textbf{r})
\end{array}\!\!\!\right)\!\!\delta(\textbf{r}-\textbf{r'}),\label{propT}\ee
where $\omega_{n}=\frac{2\pi n}{\beta}$ is the $n-$th Matsubara
frequency, and the vertex labeled $(\omega_{n},\textbf{r})$ implies
the combined sum and integral $\sum_{n=-\infty}^{\infty}\int
d^{3}r.$ The auxiliary potential is given by
$V_{J^{(0)}}(\textbf{r})=-\mu+V_{ext}(\textbf{r})+J^{(0)}(\textbf{r})$
with the functional $J^{(0)}[\Phi,\rho]$ implicitly defined by  \be
\rho(\textbf{r})= \frac{1}{2\beta}\sum _{n=-\infty}^{\infty} tr
\mathcal{G}_{J^{(0)}}(\omega_{n},\textbf{r},\textbf{r}).\label{rho0}\ee

\section{KOHN-SHAM APPROXIMATION SCHEME}\label{appr}
Now, we construct a systematic approximation scheme for bosonic fields from the expansion of the thermal DF proceeding in a way analogous to that of Valiev and Fernando~\cite{Val}. They established the Kohn-Sham approximation scheme for fermions using the effective action $\Gamma[\rho]$ with the coupling constant as the expansion parameter. In the case of bosons, the background fields do not necessarily vanish, so we have to consider the effective action $\Gamma[\Phi,\rho]$ and its expansion in powers of $\hbar$~(\ref{eafull}). 

The $K-$th order approximation at the temperature $1\over \beta$ is obtained from the thermal DF series truncated at the $K$th order
\be
\Omega^{(K)}[\Phi,\rho]=\sum_{k=0}^{K}\hbar^k
\Omega^{(k)}[\Phi,\rho].
\ee 
The approximate values of the order parameter and density are determined by the stationarity
conditions 
\be 
\frac{\delta \Omega^{(K)}}{\delta \Phi(\mathbf{r})}=
\frac{\delta \Omega^{(K)}}{\delta \Phi^{*}(\mathbf{r})}=0
\label{der1}\ee \mbox{and} \be \frac{\delta \Omega^{(K)}}{\delta
\rho(\mathbf{r})}=0,\label{der2}\ee and the chemical potential is
fixed by \be N=-\frac{\delta \Omega^{(K)}} {\delta
\mu}.
\label{chemK}
\ee

The zero-th order approximation to the thermal DF 
 \be \Omega^{(0)}[\Phi,\rho]=\int \!d^{3}r
\left[\Phi^{*}(\textbf{r})\left(
-\frac{\nabla^2}{2m}+V_{ext}(\textbf{r})-\mu\right)\Phi
(\textbf{r})+\frac{g}{2}\!
\left|\Phi(\textbf{r})\right|^{4}\right],\label{Om0}\ee yields the
time-independent Gross-Pitaevskii equation \be  \frac{\delta
\Omega^{(0)}}{\delta \Phi^{*}(\mathbf{r})}=\left(
-\frac{\nabla^2}{2m}+V_{ext}(\textbf{r})-\mu+g\!
\left|\Phi(\textbf{r})\right|^{2}\right)\Phi
(\textbf{r})=0,\label{GP} \ee and the constraint~(\ref{chemK})
takes a form \be N=-\frac{\delta \Omega^{(0)}} {\delta \mu}=\int
\!d^{3}r \left|\Phi(\textbf{r})\right|^{2}.\label{chem0}\ee This
means that in this approximation the total density is equal to the
condensate density, which is consistent with the absence of the
contributions to the particle density in the zero-th order. One has to stress that $\Omega^{(0)}[\Phi,\rho]$ does not include any temperature corrections, and can be regarded
only as an approximation to the zero-temperature DF, which describes
a full condensation into the ground state.

The first order termal DF for bosons is given by  
\ba
\Omega^{(1)}[\Phi,\rho]&=&\int
\!d^{3}r\left[\Phi^{*}(\textbf{r})\left(
-\frac{\nabla^2}{2m}+V_{ext}(\textbf{r})-\mu\right)\Phi
(\textbf{r})+\frac{g}{2}\!
\left|\Phi(\textbf{r})\right|^{4}\right]\nonumber\\&-&\hbar \int
\!d^{3}r
\rho(\textbf{r})J_{0}(\textbf{r})-\frac{\hbar}{2\beta}TrLn\mathcal{G}^{-1},\label{Om1}
\ea with
\be TrLn\mathcal{G}^{-1} =\sum_{i}\ln
\lambda_{i},\ee where the eigenvalues $\lambda_{i}$ of the operator
$\mathcal{G}^{-1}$ are determined by the Bogoliubov-de Gennes
equations
 \be
\!\!\left(\!\!\begin{array}{cc}
  -i\omega_{n}-\!\frac{\nabla^2}{2m}\!+\!2g
|\Phi(\textbf{r})|^2\!+V_{J^{(0)}}(\textbf{r})& g
\Phi^{2}(\textbf{r})\\
  g
\Phi^{*2}(\textbf{r}) &i\omega_{n}-\!\frac{\nabla^2}{2m}\!+\!2g
|\Phi(\textbf{r})|^2\!+\!V_{J^{(0)}}(\textbf{r})
\end{array}\!\!\!\right)\left(\begin{array}{c} u_{i}(\textbf{r}) \\
                           v_{i}(\textbf{r}) \\
                        \end{array}
  \right)=\lambda_{i}\left(\begin{array}{c} u_{i}(\textbf{r}) \\
                           v_{i}(\textbf{r}) \\
                        \end{array}
\right),\label{BdG}
\ee  
\be  
\frac{\delta
\Omega^{(1)}}{\delta \Phi^{*}(\mathbf{r})}=\left(
-\frac{\nabla^2}{2m}+V_{ext}(\textbf{r})-\mu+g\!
\left|\Phi(\textbf{r})\right|^{2}\right)\Phi
(\textbf{r})=0,\label{GP1} \ee
The above functional describes a system of independent particles subjected to an external potential
$V_{J^{(0)}}(\textbf{r})=-\mu+V_{ext}(\textbf{r})+J^{(0)}(\textbf{r}),$
where the function $J^{(0)}(\textbf{r})$ is unknown. Therefore, $\Omega^{(1)}[\Phi,\rho]$ can be taken as the reference system in the Kohn-Sham approximation scheme. The $K-$th order density functional may be split as
\be \Omega^{(K)}=\Omega^{(1)}+\Omega^{(K)}_{ m-b},
\ee where the many-body
contribution, $\Omega^{(K)}_{ m-b}$, contains the terms of the order
$\hbar^2$ and higher. This results in the splitting of the condition~(\ref{der2}) 
into two equations \be \frac{\delta\Omega^{(1)}}{\delta \hbar\rho
(\mathbf{r})}=-J^{(0)}(\mathbf{r})~~~~\mbox{and}
~~~~~~\frac{\delta\Omega^{(K)}_{m-b}}{\delta \hbar\rho
(\mathbf{r})}=J^{(0)}(\mathbf{r}),\label{KohnSham}\ee
where the first equality follows from the formula~(\ref{rho0}). 

The first of the above equations can be regarded as describing the
single-particle Kohn-Sham reference system. The fictitious potential
$J^{(0)}(\textbf{r})$ is determined by the second equation, which
includes many-body effects to the order $K$. An implicit character
of the relation between the density and Kohn-Sham
potential~(\ref{rho0}) leads to the self-consistent scheme for
calculating physical quantities. The equilibrium density
$\rho_{eq}^{(K)}(\textbf{r})$ and order parameter
$\Phi_{eq}^{(K)}(\textbf{r})$ have to be determined by solving
Eqs.\ref{der1}, \ref{der2} and \ref{KohnSham} self-consistently. 
The $K-$th order approximation to the grand canonical potential can
be obtained as
$\omega^{(K)}(\mu,\beta)=\Omega^{(K)}[\Phi_{eq}^{(K)},\rho_{eq}^{(K)}]$.
The Legendre construction guarantees that the density and the order
parameter determined by the exact functional $\Omega[\Phi,\rho]$ are
equal to those of the true system at the same temperature and
chemical potential. The approximation series provides
a systematic way of approaching $\rho_{eq}(\textbf{r})$,
$\Phi_{eq}(\textbf{r})$ and the grand canonical potential
$\omega(\mu,\beta)$. Approximations to other physical quantities
have to be derived first from the approximate functional
$\Gamma^{(K)}[\Phi,\rho]$, and then evaluated at
$\Phi=\Phi_{eq}^{(K)}(\textbf{r})$ and
$\rho=\rho_{eq}^{(K)}(\textbf{r})$. For example, approximations to
one-particle excitation energies may be obtained from zero-modes of the
inverse one-particle propagator \be
\Gamma^{(K)}(x,y)=\left.\frac{\delta^{2}
\Gamma^{(K)}[\Phi,\rho]}{\delta\Phi(x)
\delta\Phi^{*}(y)}\right|_{\Phi_{eq}^{(K)},{\rho_{eq}}^{(K)}} \ee
and those to density fluctuations from zero-modes of the
inverse composite propagator\be
\chi^{(K)}(x,y)=\left.\frac{\delta^{2}
\Gamma^{(K)}[\Phi,\rho]}{\delta \rho(x)\delta \rho(y)
}\right|_{\Phi_{eq}^{(K)},{\rho_{eq}}^{(K)}}. \label{dfl} \ee

\section{Conclusions}
\label{con} The Lagrangian formulation of QFT provides a rigorous
formulation of DFT for fermions and bosons. The functional
$\Gamma[\Phi,\rho]$ is defined as the effective action for both
elementary field and density operator. The formalism is universal
and can be used to study time-dependent systems and the equilibrium
phenomena. The formalism allows for extensions to other functional
theories (spin-density, current-density, ...) by introducing sources
coupled to the corresponding operators. The path integral formulation
provides a method for representing the effective action functional as
a series in powers of the Planck constant. The expansion allows to 
formulate a systematic approximation scheme, which is a generalization of the Kohn-Sham approach to bosonic fields. From an approximation to the effective action a consistent set of
approximations to physical quantities may be obtained. A single
approximation to $\Gamma[\Phi,\rho]$ describes the ground state, and
provides a way to determine approximations to other quantities, such as for example one-particle excitations or density fluctuations.

\end{document}